\documentclass[a4paper,11pt]{article}

\usepackage[latin1]{inputenc} 
\usepackage[english]{babel}
\usepackage[centerlast]{caption}
\usepackage{indentfirst}
\usepackage{amsmath,amssymb,amsbsy,latexsym,textcomp}
\usepackage{bm,bbm}
\usepackage[dvips,xdvi]{graphicx}
\usepackage{pstricks}
\usepackage{subfig}
\newtheorem{problem}{Problem}

\newcommand{\vect}[1]{%
\textup{vec}{#1}}

\newcommand{\reshape}[1]{%
{\textup{reshape}}{#1}}

\newcommand{\tr}[1]{%
{\textup{tr}}{#1}}

\newcommand{\nullspace}[1]{%
{\textup{null}}{#1}}

\newcommand{\HRule}{\rule{\linewidth}{0.5mm}}

\begin{document}

\begin{titlepage}
\begin{center}

\includegraphics[width=0.3\textwidth]{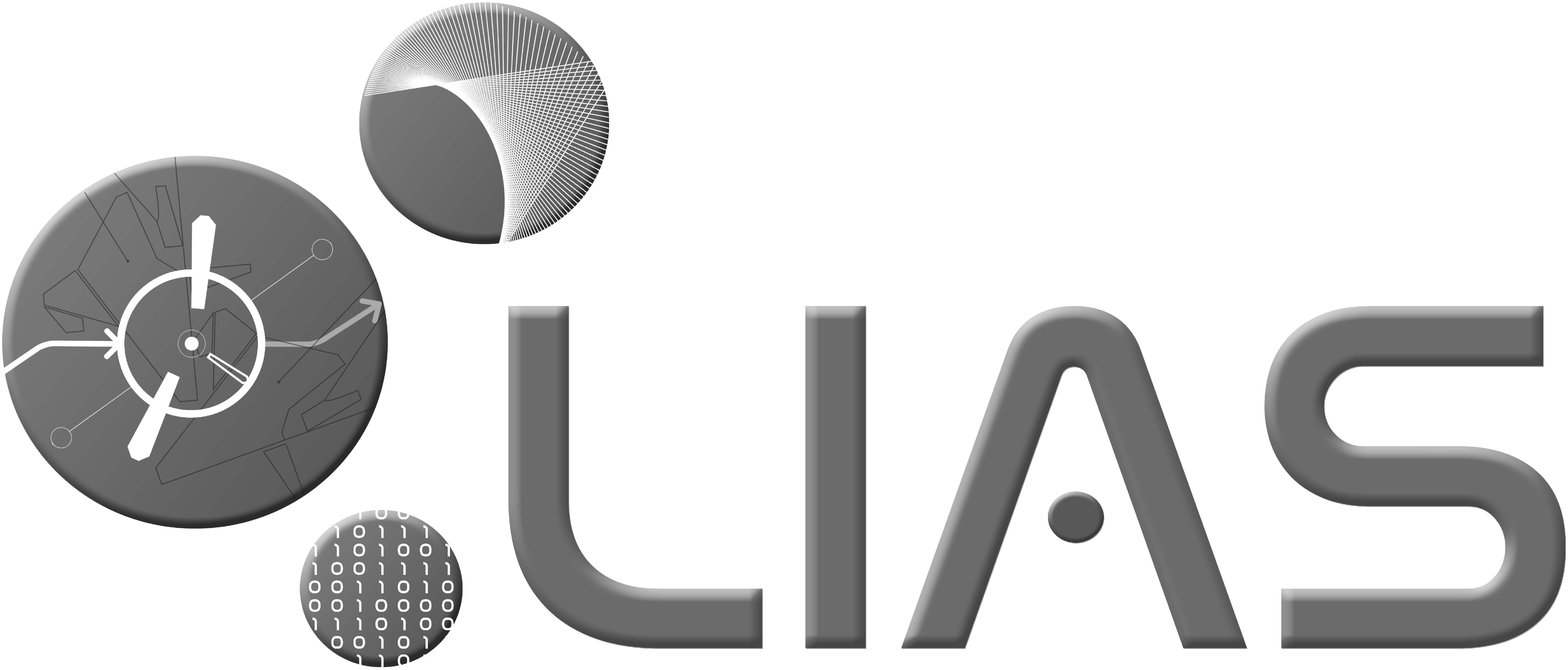} \\
\textsf{LIAS laboratory -- Poitiers University} 

\bigskip~\bigskip

\textsc{Technical Report\footnote{Technical reports from the Automatic Control group of the LIAS are available from \texttt{http://www.lias-lab.fr/publications/fulllist}}}

\bigskip

\HRule \\ \vspace{0.4cm}
\Large
\textsc{Identification of parameterized gray-box state-space systems: from a black-box linear time-invariant representation to a structured one \\ \small{Detailed derivation of the gradients involved in the cost functions}} 

\HRule

\bigskip

\normalsize
\textsf{Author:} \\
\textsc{Guillaume Mercère\footnote{LIAS Automatic Control division}},
\textsc{José Ramos\footnote{Nova Southeastern University, Fort Lauderdale, Florida, USA}},
\textsc{Olivier Prot\footnote{Limoges University, Limoges, France}}

\bigskip

\textsf{Email :} \\ \texttt{guillaume.mercere@univ-poitiers.fr} \\
\texttt{olivier.prot@unilim.fr}
\\
\texttt{jr1284@nova.edu}
\bigskip

\textsc{Report no: UP\_AS\_002}
\end{center}

\vfill
\begin{flushleft}
\textsf{Address :} \\
\textsf{Bâtiment B25 \\
2ème étage \\
2 rue Pierre Brousse \\
B.P. 633 \\
86022 Poitiers Cedex \\
web-site:} \texttt{http://www.lias-lab.fr/}
\end{flushleft}

\vfill
\begin{flushright}
\textsf{03 June 2014}
\end{flushright}

\end{titlepage}

\newpage
\thispagestyle{empty}
\mbox{}

\newpage
\setcounter{page}{1}

\title{\textsc{Identification of parameterized gray-box state-space systems: from a black-box linear time-invariant representation to a structured one\footnote{This document is available on \texttt{http://www.lias-lab.fr/perso/guillaumemercere/}} \\ \small{Detailed derivation of the gradients involved in the cost functions}}}

\author{Guillaume Mercère\thanks{Guillaume Mercère (Corresponding author) is with Poitiers University, Laboratoire d'Informatique et d'Automatique pour les Systèmes, 2 rue P. Brousse, bâtiment B25, TSA 41105, 86073 Poitiers cedex 9, France. Email:~guillaume.mercere@univ-poitiers.fr},{} Olivier Prot\thanks{Olivier Prot is with Limoges University, Institut de Recherche XLIM, 123 avenue A. Thomas, 87060 Limoges Cedex, France. Email:~olivier.prot@unilim.fr}{} and José Ramos\thanks{José Ramos is with Nova Southeastern University, Farquhar College of Arts and Sciences, Division of Mathematics, Science and Technology, 1301 College Avenue, Fort Lauderdale, FL 33314, Florida, USA. Email:~jr1284@nova.edu}}

\date{\today}

\maketitle

\begin{abstract} 
Estimating consistent parameters of a structured (gray-box) state-space representation requires a reliable initialization when the vector of parameters is computed by using a gradient-based algorithm. In the companion paper \cite{MPR}, the problem of supplying a reliable initial vector of parameters is tackled. More precisely, by assuming that a reliable initial fully-parameterized state-space model of the system is available, the paper \cite{MPR} addresses the challenging problem of transforming this initial fully-parameterized model into the structured state-space parameterization satisfied by the system to be identified. Two solutions to solve such a parameterization problem are more precisely introduced in  \cite{MPR}. First, a solution based on a null-space-based reformulation of a set of equations arising from the aforementioned similarity transformation problem is considered. Second, an algorithm dedicated to non-convex optimization is presented in order to transform the initial fully-parameterized model into the structured state-space parameterization of the system to be identified. In this technical report, a specific attention is paid to the gradient computation required by the optimization algorithms used in \cite{MPR} to solve the aforementioned problem. These gradient formulations are indeed necessary to apply the quasi-Newton Broyden-Fletcher-Goldfarb-Shanno (BFGS) methods used for the null-space based as well as the least-squares-formulated optimization techniques introduced in \cite{MPR}. For the sake of conciseness, we only focus on the smooth version of the optimization problem introduced in \cite{MPR}. Interested readers can easily extend the following results by using the chain rule as well as the sub-gradient computation techniques available in \cite{ANP08,Lew03}. 
\end{abstract}

\begin{keywords} 
System identification, parameterization, convex optimization, gray-box, black-box, null-space
\end{keywords}

\section{Introduction}

Estimating consistent parameters of a gray-box linear time-invariant state-space representation is still a difficult task in system identification. While determining the order as well as the matrices of a black-box linear state-space model is now an easy problem to solve, it is well-known that the estimated (fully-parameterized) state-space matrices are unique modulo a non-singular similarity transformation matrix. This could have serious consequences if the system being identified is a real physical system. Indeed, if the true model contains physical parameters, then the identified system could no longer have the physical parameters in a form that can be extracted easily. By assuming that the system has been identified consistently in a fully-parameterized form, the question addressed in the companion paper entitled\footnote{This paper is currently accepted as full paper for the IEEE Transactions on Automatic Control special issue for Relaxation Method in Identification and Estimation Problem.} \emph{Identification of parameterized gray-box state-space systems: from a black-box linear time-invariant representation to a structured one} then is how to recover the physical parameters from this initially estimated black-box form. Two solutions to solve such a parameterization problem are more precisely introduced in the aforecited paper. First, a solution based on a null-space-based reformulation of a set of equations arising from the aforementioned similarity transformation problem is considered. Second, an algorithm dedicated to non-convex (and non-smooth) optimization is presented to transform the initial fully-parameterized model into the structured state-space parameterization of the system to be identified. A specific constraint on the similarity transformation between both system representations is added to avoid singularity. By assuming that the physical state-space form is identifiable and the initial fully-parameterized model is consistent, it is also proved that the global solutions of these two optimization problems are unique.

In this technical report, a detailed derivation of the gradient formulas required for the optimization of the cost functions presented in \cite{MPR} is given as a supplement and background material for the interested readers. For the sake of conciseness, a specific attention is paid to the smooth versions of the non-convex optimization problems addressed in \cite{MPR}. In what follows, Section~\ref{para:problem} introduces the problem addressed in \cite{MPR} as well as the required notations to solve such a problem. Section~\ref{para:nullspace} tackles the gradient computation for the null-spaced-based technique developed in \cite{MPR}. Section~\ref{para:leastsquares} focuses on the gradient computation for the smooth least-squares formulation of Problem~\ref{prob:transf}. Finally, Section~\ref{para:concl} concludes this paper.

\section{Problem formulation and notations}\label{para:problem}

In \cite{MPR}, the authors aim at supplying solutions for the identification of structured gray-box LTI state-space systems defined by the continuous-time (CT) set of equations
\begin{subequations}\label{equ:ctsssystem}
\begin{align}
\dot{\mathbf{x}}(t) &= \bm{A}( \bm{\theta} ) \mathbf{x}(t) + \bm{B}( \bm{\theta} ) \mathbf{u}(t) \\
\mathbf{y}(t) &= \bm{C}( \bm{\theta} ) \mathbf{x}(t),
\end{align}
\end{subequations}
where $\bm{\theta}$ is a vector gathering the unknown parameters of the parameterized state-space representation $\left( \bm{A}( \bm{\theta} ),  \bm{B}( \bm{\theta} ), \bm{C}( \bm{\theta} )\right)$. More precisely, the problem of re-parameterizing a consistent fully-parameterized state-space form  $\left( \bm{\mathfrak{A}},  \bm{\mathfrak{B}}, \bm{\mathfrak{C}} \right)$ into a structured representation $\left( \mathbf{A}( \bm{\theta} ), \mathbf{B}( \bm{\theta} ), \mathbf{C}( \bm{\theta} ) \right)$ is investigated in \cite{MPR} and two solutions are suggested. Said differently, suitable solutions to the following problem are introduced in \cite{MPR}
\begin{problem}\label{prob:transf}
Consider a linear time-invariant system modeled by a minimal and structured gray-box state-space representation $\left( \bm{A}( \bm{\theta} ),  \bm{B}( \bm{\theta} ), \bm{C}( \bm{\theta} )\right)$, where the matrices are functions of relatively few unknown parameters gathered into a vector $\bm{\theta}$. Furthermore, let us assume that a consistent fully-parameterized minimal state-space realization $\left( \bm{\mathfrak{A}},  \bm{\mathfrak{B}}, \bm{\mathfrak{C}} \right)$ of the system under study is available\footnote{ The subspace-based identification techniques \cite{Lju99,VV07} are really good candidates to solve this problem. These algorithms can indeed yield consistent estimates under many different noise conditions.}. Then, the problem considered in \cite{MPR} consists in uniquely determining the similarity transformation $\mathbf{T}$ and the vector $\bm{\theta}$ satisfying
\begin{align}\label{equ:bileq}
\bm{\mathfrak{A}} \mathbf{T} &= \mathbf{T} \bm{A}( \bm{\theta} ) & \bm{\mathfrak{B}} &= \mathbf{T} \bm{B}( \bm{\theta} ) &
\bm{\mathfrak{C}} \mathbf{T} &= \bm{C}( \bm{\theta} ) .
\end{align}
\end{problem}

Two complimentary solutions are developed in \cite{MPR} to solve this problem. The first one consists in reformulating the set of equations~\eqref{equ:bileq} as the null-space problem. The second one aims at minimizing the error involved in Eq.~\eqref{equ:bileq} \cite{PL03,XL02} (see \cite{MPR} for details).
As shown in \cite{MPR}, both solutions involve specific cost functions which must be minimized to get the optimal solution of Problem~\ref{prob:transf}. As suggested in \cite{MPR}, such minimizations can be performed by using a quasi-Newton Broyden-Fletcher-Goldfarb-Shanno (BFGS) method. This BFGS method requires the computation of dedicated gradients which can be obtained analytically by resorting to standard tools of differentiations. These analytic results are given hereafter to help the user implement the algorithms introduced in \cite{MPR}.

\section{Gradient computation for the null-spaced-based technique}\label{para:nullspace}

\subsection{Objective functions for the null-space approach}

As said previously, a possible way to solve Problem~\ref{prob:transf} implies the reformulation of the set of equations~\eqref{equ:bileq} as a null-space problem. By using a standard property of the vectorization tool, \emph{i.e.}, $
  \vect{( \mathbf{M} \mathbf{N} \mathbf{P} )} = ( \mathbf{P}^\top \otimes \mathbf{M} ) \vect{( \mathbf{N} )}
$
\cite{HJ90} (where $\mathbf{M}$, $\mathbf{N}$ and $\mathbf{P}$ are matrices with compatible dimensions), it is easy to show that the set of equations~\eqref{equ:bileq} satisfies the following matrix form
\begin{equation}\label{equ:kernelprob}
\begin{split}
\underbrace{
\left[  \begin{smallmatrix}
    ( \mathbf{I}_{n_x \times n_x} \otimes \bm{\mathfrak{A}} ) & - \mathbf{I}_{n_x^2 \times n_x^2} & \mathbf{0}_{n_x^2 \times n_x n_u} & \mathbf{0}_{n_x^2 \times n_x n_y} & \mathbf{0}_{n_x^2} \\
\mathbf{0}_{n_x n_u \times n_x^2} & \mathbf{0}_{n_x n_u \times n_x^2} & \mathbf{I}_{n_x n_u \times n_x n_u} & \mathbf{0}_{n_x n_u \times n_x n_y}  & - \vect{( \bm{\mathfrak{B}} )} \\
( \mathbf{I}_{n_x \times n_x} \otimes \bm{\mathfrak{C}} ) & \mathbf{0}_{n_x n_y \times n_x^2} & \mathbf{0}_{n_x n_y \times n_x n_u} & - \mathbf{I}_{n_x n_y \times n_x n_y} & \mathbf{0}_{n_x n_y}
  \end{smallmatrix} \right]
}_{\text{matrix } \bm{\Delta} } \\
\underbrace{
  \begin{bmatrix}
    \vect{( \mathbf{T} )} \\
\vect{(\mathbf{T} \bm{A}( \bm{\theta} ))} \\
\vect{(\mathbf{T} \bm{B}( \bm{\theta} ))} \\
\vect{( \bm{C}( \bm{\theta} ) )} \\
1
  \end{bmatrix}}
_{\text{vector } \bm{\tau}} =
\mathbf{0}_{n_x^2 + n_x( n_u + n_y)},
\end{split}
\end{equation}
where the matrix $\bm{\Delta} \in \mathbb{R}^{n_x^2 + n_x( n_u + n_y) \times (2 n_x^2 + n_x( n_u + n_y) + 1)}$ is composed of known coefficients\footnote{Keep in mind that a consistent fully-parameterized state-space triplet $\left( \bm{\mathfrak{A}}, \bm{\mathfrak{B}}, \bm{\mathfrak{C}} \right)$ is assumed to be available.}, while  the vector $\bm{\tau} \in \mathbb{R}^{(2 n_x^2 + n_x( n_u + n_y) + 1)}$ gathers the unknown parameters\footnote{In order to simplify the notation, the re-definitions $n_{\tau} = 2 n_x^2 + n_x( n_u + n_y) + 1)$ and $n_{\Delta} = n_x^2 + n_x( n_u + n_y)$ will be used.}
 
As shown in \cite{MPR}, determining the optimal solution $\hat{\bm{\tau}}$ of this null-space-based problem requires following three main steps
\begin{enumerate}
\item the determination of the components of the null-space of $\bm{\Delta}$ satisfying the constraint that the last component of the estimated vector $\hat{\bm{\tau}}$ is equal to $1$,
\item the construction of a state-space form $(\bm{\mathbbm{A}}(\hat{\bm{\tau}}), \bm{\mathbbm{B}}(\hat{\bm{\tau}}), \bm{\mathbbm{C}}(\hat{\bm{\tau}}))$ and a similarity transformation matrix $\bm{\mathbbm{T}}(\hat{\bm{\tau}})$ from the estimated vector $\hat{\bm{\tau}}$, 
\item the re-structuring of the estimated matrices $(\bm{\mathbbm{A}}(\hat{\bm{\tau}}), \bm{\mathbbm{B}}(\hat{\bm{\tau}}), \bm{\mathbbm{C}}(\hat{\bm{\tau}}))$ so that the structural constraints satisfied by $\bm{A}( \bm{\theta} )$, $\bm{B}( \bm{\theta} )$ and $\bm{C}( \bm{\theta} )$ are verified.
\end{enumerate}
As shown in \cite{MPR}, the last phase, which consists in re-structuring the matrices $(\bm{\mathbbm{A}}(\hat{\bm{\tau}}), \bm{\mathbbm{B}}(\hat{\bm{\tau}}), \bm{\mathbbm{C}}(\hat{\bm{\tau}}))$, requires a specific optimization algorithm. In this technical report, we only focus on this final re-structuring step. It is indeed the only one which entails specific gradient computations. The interested reader should read \cite{MPR} to get more details concerning the complimentary steps composing this null-space-based technique.

Mathematically, constraining the data set $( \bm{\mathbbm{A}}(\bm{\tau}), \bm{\mathbbm{B}}(\bm{\tau}), \bm{\mathbbm{C}}(\bm{\tau}))$ in order to satisfy specific structural constraints known from the parameterization  $\bm{A}( \bm{\theta} )$, $\bm{B}( \bm{\theta} )$ and $\bm{C}( \bm{\theta} )$ can be modelled with the help of a function $f_S$ defined as follows
\begin{equation}
\begin{matrix}
  f_S: & \mathbb{R}^{n_{\Delta}} & \longrightarrow & \mathbb{R}^+ \\
& \bm{\gamma} & \longmapsto & f_S(\bm{\gamma})
\end{matrix}
\end{equation}
where, for $\bm{\tau} \in \mathbb{R}^{n_\tau}$ and
\begin{equation}
  \bm{\gamma}(\bm{\tau}) =
\begin{bmatrix}
\vect{( \bm{\mathbbm{A}}(\bm{\tau}) )} \\
\vect{( \bm{\mathbbm{B}}(\bm{\tau}) )} \\
\vect{( \bm{\mathbbm{C}}(\bm{\tau}) )}
\end{bmatrix} \in \mathbb{R}^{n_{\Delta}}
\end{equation}
such that
\begin{itemize}
\item $f_S: \mathbb{R}^{n_{\Delta}} \mapsto \mathbb{R}^+$ is a continuous function,
\item if $f_S(\bm{\gamma}) = 0$, then $\bm{\gamma}$ satisfies the aforementioned parameterization constraints.
\end{itemize}
Thus, $\tau$ can be identified by solving
\begin{equation}\label{equ:optpb}
\min_{\bm{\tau} \in \mathcal{X} \setminus \mathcal{S}} f_S(\bm{\gamma}(\bm{\tau}))
\end{equation}
where\footnote{see \cite{MPR} for a definition of $\mathcal{S}$.} $\mathcal{X}$ is the affine space of admissible $\bm{\tau}$ vectors satisfying $\bm{\tau}(end) = 1$, \emph{i.e.},
$
  \mathcal{X} = \left\{ \bm{\tau} \in \nullspace(\bm{\Delta}) : \bm{\tau}(end) = 1 \right\} 
$
and where $\bm{\tau} \longmapsto \gamma\left( \bm{\tau} \right)$ is a rational function.

When  $\bm{A}(\bm{\theta})$, $\bm{B}(\bm{\theta})$ and $\bm{C}(\bm{\theta})$ are affine functions of the parameter vector $\bm{\theta}$, \emph{i.e.}, when 
\begin{equation}
\bm{\kappa}(\bm{\theta}) =
\begin{bmatrix}
\vect(\bm{A}(\bm{\theta})) \\
\vect(\bm{B}(\bm{\theta})) \\
\vect(\bm{C}(\bm{\theta}))
\end{bmatrix} = \bm{\kappa}_0 + \mathbf{K} \bm{\theta}
\end{equation}
where $\bm{\kappa}_0 \in \mathbb{R}^{n_{\Delta}}$ and $\mathbf{K} \in \mathbb{R}^{n_{\Delta} \times n_{\theta}}$, a convenient choice for the $f_S$ function may be
\begin{equation}
  f_S(\bm{\gamma}) = \inf_{\bm{\theta} \in \mathbb{R}^{n_{\theta}}} \left\| \bm{\kappa}_0 + \mathbf{K} \bm{\theta} - \bm{\gamma} \right\|_2^2.
\end{equation}
Computing the projection of $\bm{\gamma} \in \mathbb{R}^{n_{\Delta}}$ onto the affine space $\left\{ \bm{\kappa}(\bm{\theta}) : \bm{\theta} \in \mathbb{R}^{n_{\bm{\theta}}} \right\}$ can be performed by using a singular value decomposition of $\mathbf{K}$. Indeed, this tool leads to the solution
\begin{equation}
  \bm{\theta}^* = (\mathbf{K}^\top \mathbf{K})^\dag \mathbf{K}^\top (\bm{\kappa}_0 -\bm{\gamma} )
\end{equation}
where $(\bullet)^\dag$ is the Moore-Penrose pseudo inverse \cite{HJ90} computed from the SVD of $\mathbf{K}$. By using this solution explicitly, the criterion $f_S(\bm{\gamma})$ becomes
\begin{equation}
  f_S(\bm{\gamma}) = \left\| \mathbf{M}_{\mathbf{K}}(\bm{\kappa}_0 - \bm{\gamma} ) \right\|_2^2
\end{equation}
where the orthogonal projection $\mathbf{M}_{\mathbf{K}}$ 
satisfies
$
  \mathbf{M}_{\mathbf{K}} = \mathbf{K} (\mathbf{K}^\top \mathbf{K})^\dag \mathbf{K}^\top - \mathbf{I}_{n_{\Delta} \times n_{\Delta}} .
$
This simplification leads to a much easier optimization problem
\begin{equation}\label{equ:optpb2}
  \min_{\bm{\tau} \in \mathcal{X} \setminus \mathcal{S}} \left\| \mathbf{M}_{\mathbf{K}} (\bm{\kappa}_0 - \bm{\gamma}(\bm{\tau}) ) \right\|_2^2 .
\end{equation}
In practice, the vectors and matrices $\bm{\kappa}_0$, $\mathbf{K}$ and $\mathbf{M}_{\mathbf{K}}$ can be computed from the structure of $(\bm{A}(\bm{\theta}),\bm{B}(\bm{\theta}),\bm{C}(\bm{\theta}))$ by using computer algebra.

\subsection{Parameterization of the affine space $\mathcal{X}$}

In practice, a parameterization of the affine space $\mathcal{X}$ can be used to solve the aforementioned optimization problem. This parameterization is based on a three-step procedure. First, a basis of the null-space of $\bm{\Delta}$ is computed by using an SVD. By denoting this basis by $\mathbf{Z} \in \mathbb{R}^{n_{\tau} \times n_{Z}}$, where $n_Z = \dim{(\nullspace(\bm{\Delta}))}$, the second step consists in determining\footnote{This can be done, \emph{e.g.}, by generating $\bm{\beta}_0$ randomly, then by computing $\mathbf{Z} \bm{\beta}_0$ and finally by fixing $\bm{\beta}_0 = \bm{\beta}_0/((\mathbf{Z} \bm{\beta}_0)(end))$, where $(\mathbf{Z} \bm{\beta}_0)(end)$ denotes the last component of the vector $\mathbf{Z} \bm{\beta}_0$.} a vector $\bm{\beta}_0 \in \mathbb{R}^{n_{Z}}$ such that $\mathbf{Z} \bm{\beta}_0 \in \mathcal{X}$, \emph{i.e.}, $\mathbf{Z} \bm{\beta}_0$ satisfies the constraint that the last component of $\bm{\tau}$ equals $1$. By having access to this vector $\bm{\beta}_0$, the third step aims at computing a basis of the null-space of the last row of $\mathbf{Z}$. By doing so, a matrix $\mathbf{Z}_2 \in \mathbb{R}^{n_{Z} \times (n_{Z}-1)}$ can be built such that, for all $\bm{\alpha} \in  \mathbb{R}^{n_{Z}-1}$, the last component of the vector $\mathbf{Z} \mathbf{Z}_2 \bm{\alpha}$ is zero. By using these three steps, the affine space $\mathcal{X}$ can be parameterized as follows
\begin{equation}\label{equ:Xparam}
  \mathcal{X} = \left\{ \mathbf{Z} (\bm{\beta}_0 + \mathbf{Z}_2 \bm{\alpha}) : \bm{\alpha} \in  \mathbb{R}^{n_{Z}-1} \right\} .
\end{equation}
Such a parameterization implies that $n_{\mathcal{X}} = n_{Z}-1$. By using Eq.~\eqref{equ:Xparam}, the optimization problem~\eqref{equ:optpb} becomes
\begin{equation}
  \min_{\bm{\alpha} \in \mathbb{R}^{n_{\mathcal{X}}}} f_S( \bm{\gamma}(\mathbf{Z} (\bm{\beta}_0 + \mathbf{Z}_2 \bm{\alpha})) ) .
\end{equation}
By extension, the optimization problem~\eqref{equ:optpb2} satisfies
\begin{equation}\label{equ:optpb3}
  \min_{\bm{\alpha} \in \mathbb{R}^{n_{\mathcal{X}}}} \left\| \mathbf{M}_{\mathbf{K}} (\bm{\kappa}_0 - \bm{\gamma}(\mathbf{Z} (\bm{\beta}_0 + \mathbf{Z}_2 \bm{\alpha}))) \right\|_2^2 .
\end{equation}

\subsection{Gradient formulation}

We now derive the expression of the gradient of cost function $h$
defined by
\begin{equation}
\begin{matrix}
  h: & \mathbb{R}^{n_{\tau}} & \longrightarrow & \mathbb{R}^+ \\
& \bm{\tau} & \longmapsto & \left\| \mathbf{M}_{\mathbf{K}} (\bm{\kappa}_0 - \bm{\gamma}(\bm{\tau}) ) \right\|_2^2 .
\end{matrix}
\end{equation}
Clearly, the function $h$ can be written as a composite function, \emph{i.e.}, $\bm{\tau} \longmapsto f_S(\gamma(\bm{\tau}))$. Thus, the chain rule can be applied to derive its gradient. First, since the function $f_S(\bullet)$ is a quadratic function, it holds that
\begin{equation}
\nabla_{\bm{\gamma}} f_S(\gamma) = - 2 \mathbf{M}_{\mathbf{K}}^\top \mathbf{M}_{\mathbf{K}}
(\bm{\kappa}_0 - \bm{\gamma} ).
\end{equation}
Second, the Jacobian of the functions
\begin{subequations}
  \begin{align}
    &\begin{matrix}
      \mathbb{R}^{n_{\tau}} & \longrightarrow & \mathbb{R}^{n_x^2} \\
      \bm{\tau} & \longmapsto & \vect(\mathbbm{A}(\bm{\tau}))
    \end{matrix} \\
    &\begin{matrix}
      \mathbb{R}^{n_{\tau}} & \longrightarrow & \mathbb{R}^{n_x n_u} \\
      \bm{\tau} & \longmapsto & \vect(\mathbbm{B}(\bm{\tau}))
    \end{matrix}\\
    &\begin{matrix}
      \mathbb{R}^{n_{\tau}} & \longrightarrow & \mathbb{R}^{n_x n_y} \\
      \bm{\tau} & \longmapsto & \vect(\mathbbm{C}(\bm{\tau}))
    \end{matrix}
\end{align}
  \end{subequations}
 are respectively
 \begin{subequations}
   \begin{align}
     \mathbf{J}_{\mathbbm{A}}(\bm{\tau}) = & - \left(\left(\mathbbm{T}(\bm{\tau})^{-1} \mathbbm{A}(\bm{\tau})\right)^\top \otimes \mathbbm{T}(\bm{\tau})^{-1}\right) \mathbf{P}_{\mathbbm{T}} + \left(I_{n_x} \otimes \mathbbm{T}(\bm{\tau})^{-1}\right) \mathbf{P}_{\mathbbm{A}} \\
     \mathbf{J}_{\mathbbm{B}}(\bm{\tau}) = & - \left(\left(\mathbbm{T}(\bm{\tau})^{-1} \mathbbm{B}(\bm{\tau})\right)^\top \otimes \mathbbm{T}(\bm{\tau})^{-1}\right) \mathbf{P}_{\mathbbm{T}} + \left(I_{n_u} \otimes \mathbbm{T}(\bm{\tau})^{-1}\right) \mathbf{P}_{\mathbbm{B}} \\
     \mathbf{J}_{\mathbbm{C}}(\bm{\tau}) = & \mathbf{P}_{\mathbbm{C}}
   \end{align}
 \end{subequations}
where the selection matrices $\mathbf{P}_{\bullet}$ are defined as follows
\begin{subequations}
  \begin{align}
    \mathbf{P}_{\mathbbm{T}} &=
\begin{bmatrix}
\mathbf{I}_{n_x^2 \times n_x^2} & \mathbf{0}_{n_x^2 \times n_x^2 + n_x(n_u + n_y) +1 }
\end{bmatrix} \\
\mathbf{P}_{\mathbbm{A}} &=
\begin{bmatrix}
\mathbf{0}_{n_x^2 \times n_x^2} & \mathbf{I}_{n_x^2 \times n_x^2} & \mathbf{0}_{n_x^2 \times  n_x(n_u + n_y) +1 }
\end{bmatrix} \\
\mathbf{P}_{\mathbbm{B}} &=
\begin{bmatrix}
\mathbf{0}_{n_x n_u \times 2 n_x^2} & \mathbf{I}_{n_x n_u \times n_x n_u} & \mathbf{0}_{n_x n_u \times  n_x n_y +1 }
\end{bmatrix} \\
\mathbf{P}_{\mathbbm{C}} &=
\begin{bmatrix}
\mathbf{0}_{n_x n_y \times 2 n_x^2 + n_x n_u} & \mathbf{I}_{n_x n_y \times n_x n_y} & \mathbf{0}_{n_x n_y \times 1 }
\end{bmatrix},
  \end{align}
\end{subequations}
and where $\reshape{(\bullet,n_1, n_2)}$ returns the $n_1 \times n_2$ matrix whose elements are taken columnwise from $\bullet$. 

By using the chain rule, we are now able to derive the gradient formulation for cost function $h$
\begin{equation}
\nabla_{\bm{\tau}} h(\bm{\tau}) =
- 2 \begin{bmatrix}
\mathbf{J}_{\mathbbm{A}}^\top(\bm{\tau}) & \mathbf{J}_{\mathbbm{B}}^\top(\bm{\tau}) & \mathbf{J}_{\mathbbm{C}}^\top(\bm{\tau}) 
\end{bmatrix}
\mathbf{M}_{\mathbf{K}}^\top \mathbf{M}_{\mathbf{K}}
(\bm{\kappa}_0 - \bm{\gamma}(\bm{\tau}) ).
\end{equation}
In practice, it is more convenient to optimize the cost function~\eqref{equ:optpb3} than the criterion~\eqref{equ:optpb2}. This problem requires applying (again) the chain rule in order to
compute the gradient of function 
\begin{equation}
  \begin{matrix}
    \bar{h}: &  \mathbb{R}^{n_{Z}-1} & \longrightarrow & \mathbb{R}^{+} \\
      & \bm{\alpha} & \longmapsto & \left\| \mathbf{M}_{\mathbf{K}} (\bm{\kappa}_0 - \bm{\gamma}(\mathbf{Z} (\bm{\beta}_0 + \mathbf{Z}_2 \bm{\alpha}))) \right\|_2^2.
    \end{matrix}
\end{equation}
After straightforward calculations, we get finally
\begin{multline}
\nabla_{\bm{\alpha}} \bar{h} = 
- 2 \, \mathbf{Z}_2^\top \mathbf{Z}^\top \begin{bmatrix}
\mathbf{J}_{\mathbbm{A}}^\top(\bm{\tau}) & \mathbf{J}_{\mathbbm{B}}^\top(\bm{\tau}) & \mathbf{J}_{\mathbbm{C}}^\top(\bm{\tau}) 
\end{bmatrix} \\
\mathbf{M}_{\mathbf{K}}^\top \mathbf{M}_{\mathbf{K}}
(\bm{\kappa}_0 - \bm{\gamma}\left( \mathbf{Z} (\bm{\beta}_0 + \mathbf{Z}_2
\bm{\alpha}\right)) ).
\end{multline}

\section{Gradient computation for the least-squares formulation of Problem~\ref{prob:transf}}\label{para:leastsquares}

A second solution to solve Problem~\ref{prob:transf} consists in  resorting to a least-squares formulation of Eq.~\eqref{equ:bileq}, \emph{i.e.}, the optimization of a cost function\footnote{Notice that, in \cite{MPR}, this cost function is modified to allow for the condition number of the involved similarity transformation. Under such constraint, a non-smooth optimization problem must be tackled. This is not the case herein because, in this technical report, we only focus on the smooth version of the optimization problem introduced in \cite{MPR}.} $\digamma(\bm{\theta},\mathbf{T})$ defined as follows \cite{PL03,XL02}
\begin{equation}
  \label{equ:ABCcriterion}
  \digamma ( \bm{\theta}, \mathbf{T} ) = \left\| \bm{\mathfrak{A}} \mathbf{T} -  \mathbf{T} \bm{A}( \bm{\theta} ) \right\|^2_F 
+ \left\| \bm{\mathfrak{B}} - \mathbf{T} \bm{B}( \bm{\theta}) \right\|^2_F + \left\| \bm{\mathfrak{C}} \mathbf{T} - \bm{C}( \bm{\theta} ) \right\|^2_F 
\end{equation}
where $\left\| \bullet  \right\|^2_F$ is the Frobenius norm \cite{HJ90}. 

In order to perform the estimation of the vector $\bm{\theta}$, as well as the similarity transformation matrix $\mathbf{T}$, \emph{i.e.}, in order to find a local optimum of the cost function $\digamma$, a quasi-Newton Broyden-Fletcher-Goldfarb-Shanno (BFGS) method can be used \cite{NW06}. As said previously, to set up this method, the gradients of the cost function $\digamma$ with respect to $\bm{\theta}$ and $\mathbf{T}$ must be computed. This gradient computation can be performed as follows. By assuming that the structured state-space matrices depend on $\bm{\theta}$ in an affine manner, the vectorized version of the matrices $\bm{A}(\bm{\theta})$, $\bm{B}(\bm{\theta})$ and $\bm{C}(\bm{\theta})$ can be defined as follows
\begin{subequations}
  \begin{align}
    \vect{ \left( \bm{A}( \bm{\theta} ) \right) } &= \bm{\kappa}_A + \mathbf{K}_A \bm{\theta} \\ \vect{ \left( \bm{B}( \bm{\theta} )\right) } &= \bm{\kappa}_B + \mathbf{K}_B \bm{\theta} \\ \vect{ \left( \bm{C}( \bm{\theta} ) \right)
    } &= \bm{\kappa}_C + \mathbf{K}_C \bm{\theta} 
  \end{align}
\end{subequations}
where $\bm{\kappa}_\bullet$ and $\mathbf{K}_\bullet$ are constant vectors and matrices of appropriate dimensions respectively.
Then, the expression of the gradient of $\digamma$ with respect to $\bm{\theta}$ is 
\begin{multline}\label{equ:gradientlinear}
\nabla_{\bm{\theta}} \digamma \left( \bm{\theta}, \mathbf{T} \right) = - 2 \vect{ \left( \mathbf{T}^\top \bm{\mathfrak{A}} \mathbf{T} - \mathbf{T}^\top \mathbf{T} \bm{A}( \bm{\theta} ) \right) } \mathbf{K}_A \\
- 2 \vect{ \left( \mathbf{T}^\top \bm{\mathfrak{B}} - \mathbf{T}^\top \mathbf{T} \bm{B}( \bm{\theta} ) \right) } \mathbf{K}_B -2
 \vect{ \left( \bm{\mathfrak{C}} \mathbf{T} - \bm{C}( \bm{\theta} ) \right) } \mathbf{K}_C .
\end{multline}
It is important to point out that Eq.~\eqref{equ:gradientlinear} remains valid even in the non-linear case by substituting the matrices $\mathbf{K}_A$, $\mathbf{K}_B$ and $\mathbf{K}_C$ for the Jacobian matrices of $\vect{ \left( \bm{A}( \bm{\theta} ) \right) }$, $\vect{ \left( \bm{B}( \bm{\theta} ) \right) }$ and $\vect{ \left( \bm{C}( \bm{\theta} ) \right) }$ with respect to $\bm{\theta}$. The gradient of the cost function $\digamma$ with respect to $\mathbf{T}$ can be derived by using the fact that $\|{\bullet}\|_F^2 = \tr{(\bullet^\top \bullet)}$ \cite{HJ90}, \emph{i.e.},
\begin{multline}
\nabla_{\mathbf{T}} \digamma \left( \bm{\theta}, \mathbf{T} \right) = 2 \left( \bm{\mathfrak{C}}^\top \bm{\mathfrak{C}} \mathbf{T} -  \bm{\mathfrak{C}}^\top \bm{C}( \bm{\theta} ) \right) 
 \\ + 2 \left( \bm{\mathfrak{A}}^\top \bm{\mathfrak{A}} \mathbf{T} - \bm{\mathfrak{A}}^\top \mathbf{T} \bm{A}( \bm{\theta} ) - \bm{\mathfrak{A}} \mathbf{T} \bm{A}^\top( \bm{\theta} ) + \mathbf{T} \bm{A}( \bm{\theta} ) \bm{A}^\top( \bm{\theta} ) \right) \\ + 2 \left( \mathbf{T} \bm{B}( \bm{\theta}) \bm{B}^\top( \bm{\theta}) - \bm{\mathfrak{B}} \bm{B}^\top( \bm{\theta}) \right) .
\end{multline}

\section{Conclusion}\label{para:concl}

In this technical report, a specific attention has been paid to the gradient computation required by the optimization algorithms involved to solve Problem~\ref{prob:transf} tackled in \cite{MPR}. These gradient formulations are indeed necessary to apply the quasi-Newton Broyden-Fletcher-Goldfarb-Shanno (BFGS) methods used for the null-space based as well as the least-squares-formulated optimization techniques introduced in \cite{MPR}. Notice that, in \cite{MPR}, the problem of constraining the condition number of the involved similarity transformation matrix is also addressed. Because such a constraint involves using the maximum eigenvalue function and, by extension, leads to non-smooth cost functions, sub-gradient computations are necessary as well. This non-smooth case has not been addressed in this technical report. However, these sub-gradients can be computed by applying the chain rule as performed previously. See \cite{ANP08,Lew03} for more details about the sub-gradient computation of the maximum eigenvalue function.

\bibliographystyle{plain}

\begin{thebibliography}{1}

\bibitem{ANP08}
P.~Apkarian, D.~Noll, and O.~Prot.
\newblock A trust region spectral bundle method for non-convex eigenvalue
  optimization.
\newblock {\em {SIAM} Journal of Optimization}, 19:281--306, 2008.

\bibitem{HJ90}
R.~Horn and C.~Johnson.
\newblock {\em Matrix analysis}.
\newblock Cambridge University Press, 1990.

\bibitem{Lew03}
A.~Lewis.
\newblock The mathematics of eigenvalue optimization.
\newblock {\em Mathematical Programming}, 97:155--176, 2003.

\bibitem{Lju99}
L.~Ljung.
\newblock {\em System identification. Theory for the user}.
\newblock Prentice Hall, Upper Saddle River, 2nd edition, 1999.

\bibitem{MPR}
G.~Merc{\`e}re, O.~Prot, and J.~Ramos.
\newblock Identification of parameterized gray-box state-space systems: from a
  black-box linear time-invariant representation to a structured one.
\newblock {IEEE} Transactions on Automatic Control, accepted for publication,
  2014.

\bibitem{NW06}
J.~Nocedal and S.~Wright.
\newblock {\em Numerical Optimization}.
\newblock Springer-Verlag, 2006.

\bibitem{PL03}
P.~Parrilo and L.~Ljung.
\newblock Initialization of physical parameter estimates.
\newblock In {\em Proceedings of the {IFAC} Symposium on System
  Identification}, Rotterdam, The Netherlands, August 2003.

\bibitem{VV07}
M.~Verhaegen and V.~Verdult.
\newblock {\em Filtering and system identification: a least squares approach}.
\newblock Cambridge University Press, 2007.

\bibitem{XL02}
L.~Xie and L.~Ljung.
\newblock Estimate physical parameters by black box modeling.
\newblock In {\em Proceedings of the Chinese Control Conference}, Hangzhou,
  China, June 2002.

\end{thebibliography}

\end{document}